\documentstyle[11pt,fps]{article}
\newcommand{\Z}{Z \!\!\! Z}

\newcommand{\eps}{\epsilon}
\newcommand{\gf}{\gamma_{5}}
\newcommand{\vare}{\varepsilon}

\begin{document}

\begin{center}
{\large\bf Solutions of the Ginsparg-Wilson relation\\
\ \\
and improved domain wall fermions}
\end{center}
\ \\
\begin{center}
W. Bietenholz \\
\ \\
HLRZ c/o Forschungszentrum J\"{u}lich, D-52425 J\"{u}lich, Germany \\
\ \\
Preprint HLRZ 1998-17, March 1998
\end{center}
\ \\
{\it We discuss a number of lattice fermion actions 
solving the Ginsparg-Wilson relation.
We also consider short ranged approximate solutions.
In particular, we are interested in reducing the
lattice artifacts, while avoiding (or suppressing) additive
mass renormalization. In this context, we also arrive at
a formulation of improved domain wall fermions.} \\

{\bf I. The remnant chiral symmetry of fermionic fixed point actions} \\

It is a notorious problem to find a lattice regularization
for fermions, which preserves chiral symmetry and avoids
species doubling. The celebrated
Nielsen Ninomiya No-Go theorem excludes the reproduction
of the chiral symmetry manifestly in a local lattice action
of a single fermion,
if some plausible assumptions are respected \cite{NN}
(here, locality means that
the inverse propagator is analytic in momentum space).
However, the manifest chiral invariance of the lattice action is not
badly needed to its full extent: with respect to important issues,
it is sufficient to preserve just a remnant
chiral symmetry, circumventing the Nielsen Ninomiya theorem.\\

One way to tackle the problem uses the technique of
block variable renormalization group transformations (RGTs).
Under such transformations, the partition functions and all
expectation values -- hence the physical contents of a theory --
remain invariant \cite{WilKog}. Usually one
maps the theory from some fine lattice to a coarse one,
increasing the lattice spacing by a factor $n$.
An infinite number of iterations at infinite correlation length
leads -- for suitable RGT parameters -- to a fixed point action (FPA)
of the considered RGT.
If we send the blocking factor $n\to \infty$, 
and we refer to coarse lattice
units from the beginning, then we map a continuum theory
on the lattice, without altering its physical contents.
This mapping, which generates the FPA in just one step,
involves a functional integral over the
continuum fields.

For free fermions, such a blocking from the continuum can
be performed by the RGT
\begin{eqnarray} \label{RGT}
&& \hspace{-10mm} e^{-S[\bar \Psi , \Psi ]} = \int D\bar \psi D \psi \
e^{ - s[ \bar \psi , \psi ]}
\times \\
&&  \hspace{-10mm} 
\exp \Big\{ \! - \!\!\! \sum_{x,y\in \Z^{d}} 
\Big[ \bar \Psi_{x} - \int dr \Pi (x-r)\bar \psi (r)\Big] \
\alpha_{x,y}^{-1} \ \Big[ \Psi_{y} - \int dr' \Pi (y-r') \psi (r')
\Big] \Big\} , \nonumber
\end{eqnarray}
where $\bar \psi , \ \psi$ are the continuum fields,
$\bar \Psi , \ \Psi $ the lattice fields, $s$ is the continuum
action, and $S$ the fixed point lattice action.
The matrix $\alpha$ and the convolution function $\Pi$
specify the RGT (the latter is peaked
around 0 and normalized as $\int dr \Pi (r)=1$).
For a massless fermion, we obtain
in momentum space the fixed point propagator \cite{old,UJW}
\begin{equation} \label{prop}
G(p) = \sum_{l \in \Z^{d}} \frac{\Pi^{2} (p+2\pi l)}
{i (p_{\mu}+2\pi l_{\mu})\gamma_{\mu}} + \alpha (p)\ , \quad
p \in B = ]-\pi ,\pi ]^{d} .
\end{equation}
It turns out that the FPA characterized by this
propagator is free of fermion doubling \cite{GW,UJW}.

In the limit $\alpha = 0$, the blocking of the fermion
fields is implemented by $\delta$ functions. 
In this case, and generally for
{\em any RGT with} $\{ \alpha , \gamma_{5} \} =0$,
the lattice action is chirally invariant
($\{ , \}$ denotes the anticommutator).
There is no contradiction with the Nielsen Ninomiya theorem, 
because in all these cases the action is {\em non-local} \cite{GW}:  
$G^{-1}(p)$ has poles at the corners of the Brillouin zone $B$.
This is a non-locality of the type proposed by Rebbi \cite{Rebbi}.
His fermion proposal was dismissed, however, because it implied 
zero axial anomaly \cite{antiReb},
and -- as an extension of the Nielsen Ninomiya theorem --
it has been shown that this is the case for a whole class of
non-local fermions with poles in $G^{-1}(p)$ \cite{Peli}. 
However, a consistent blocking of the fermionic fields, 
of the free gauge fields, of the interaction term to 
the first order in the gauge coupling, and of the axial charge and
current, {\em does} reproduce the axial anomaly correctly on 
the lattice \cite{BW}.
\footnote{The reason why the generalized No-Go theorem does not 
apply is that the consistent perfect axial lattice current is 
also non-local, whereas it has been assumed to be local in the 
proof of Ref. \cite{Peli}.}

Now we turn to the case
where the action (and the axial current) becomes {\em local},
so we now deal with $\{ \alpha ,\gamma_{5} \} \neq 0$. 
In this case, which is more interesting in view of practical 
applications, the RGT term is not chirally invariant, and therefore 
the chiral symmetry in its naive form is not manifest in the fixed point
action. Hence the No-Go theorem does not apply, but the
chiral symmetry is still represented correctly {\em in the observables},
due to the very nature of the RGT \cite{BW}. This is an elegant way to 
by-pass the No-Go theorem.

Since integrating out the continuum fields requires 
a functional integral, the FPA -- and more generally,
the perfect action -- can be made explicit in perturbation theory.
Beyond that,
also non-perturbative properties, which are known in the continuum, 
can be reproduced correctly on the lattice, but the expressions 
for the lattice action seem to become somewhat symbolic.
However, it is very interesting that the 
Atyiah-Singer index theorem still holds even for the classically
perfect gauge action \cite{HLN} (used together with a classically perfect 
topological charge \cite{topo}). Still, the action described in this way
(in terms of classical inverse blocking by minimization) 
is somehow implicit, but certainly not symbolic any more.
In the Schwinger model, it was indeed possible to confirm
numerically the index theorem for a classically perfect
action (to a good accuracy) \cite{FaLa}.

The fixed point propagator (\ref{prop}) obeys
\begin{equation} \label{GW1}
\{ G , \gamma_{5} \} = \{ \alpha , \gamma_{5} \} .
\end{equation}
In the case of a local -- but not chirally symmetric --
term $\alpha$, this can be viewed as a remnant chiral symmetry
of the lattice action. This relation has been postulated by
Ginsparg and Wilson as the `softest way' to break chiral
invariance in the lattice action \cite{GW}.
In this form, the remnant chiral symmetry is particularly transparent:
it is inherent to the propagator up to a {\em local violation}.
\footnote{In contrast, the usual chiral symmetry breaking terms like 
mass or a Wilson term correspond to a local violation in the {\em inverse} 
propagator. There the symmetry breaking in the propagator
becomes non-local. The same happens if we block a massive fermion
from the continuum (the mass is added in the denominator
of the propagator (\ref{prop}) \cite{QuaGlu}).}
In particular, it was demonstrated in Ref. \cite{GW}
that the Ginsparg-Wilson relation (GWR), eq. (\ref{GW1}) 
with local anticommutators, is sufficient to obtain the correct
triangle anomaly, and 
also the soft pion
theorems were expected to be reproduced correctly. Furthermore,
this property is sufficient to obtain the correct anomaly of
$ Tr (\gamma_{5}G^{-1})$ \cite{HLN,ML}. In addition, it has been shown
that eq. (\ref{GW1}) implies a continuous symmetry, 
\footnote{Blocking a SUSY theory from the continuum could reveal an
analogous remnant lattice supersymmetry
(but the resulting action differs from the one
proposed in Ref. \cite{SUSY}).}
since the action is invariant under the substitution \cite{ML}
\begin{equation} \label{trafo}
\bar \Psi \to \bar \Psi (1 + \eps 
[ 1 - G^{-1} \alpha ] \gamma_{5} ) \ , \
\Psi \to (1 + \eps \gamma_{5} [1 - \alpha G^{-1}]) \Psi
\end{equation}
to $O(\eps )$, as we see if we write the GWR in the form
\begin{equation} \label{GW2}
\{ \gamma_{5}, G^{-1} \} = G^{-1} \{ \gamma_{5},\alpha \} G^{-1} \ .
\end{equation}
All this sheds light on the
``miraculous way'' the FPA finds to circumvent the No-Go theorem.

Actually, the transformation (\ref{trafo}) is a
generalization of the case $\alpha = 1/2$ considered in
Ref. \cite{ML}, to any $\alpha$ permissible in the GWR.
However, in the following we also
put special emphasis on $\alpha = 1/2$. 
If we perform the blocking (\ref{RGT}) with the
standard block average scheme ($\Pi (u) = 1$ if $\vert u_{\mu} \vert
\leq 1/2$, $\mu =1\dots d$, and $\Pi(u)=0$ otherwise),
then the choice $\alpha =1 /2 $ (or $\alpha = -1/2$) 
does in fact optimize the locality
of the free FPA, which we write in coordinate space as
\begin{equation} \label{act}
S [\bar \Psi ,\Psi ] = \sum_{x,r\in \Z^{d}} \bar \Psi_{x}
[\rho_{\mu}(r)\gamma_{\mu} + \lambda (r)] \Psi_{x+r} \ .
\end{equation}
For this choice,
the couplings $\rho_{\mu}(r)$, $\lambda(r)$ are restricted 
to nearest neighbors in $d=1$, and in $d>1$ their exponential 
decay is extremely fast.
\footnote{For instance, on the 4-space diagonal the decay
behavior is $\rho_{\mu}(n,n,n,n) \propto \exp (-4.94 \cdot n)$,
$\lambda( n,n,n,n) \propto \exp (-4.97 \cdot n)$.} 
The numeric values of the leading couplings in $d=4$ are
given in Ref. \cite{QuaGlu}.

If we consider the form (\ref{act}) as a general ansatz for the
lattice action, normalized so that $\sum_{r} r_{\mu}\rho_{\mu}(r)
=1/2$ ({\em no} sum over $\mu$)
and $\sum_{r} \lambda (r) = 0$ (which means that the fermion
is massless), then we see that it is hardly
possible to find an ultralocal solution of the GWR
(by ``ultralocal'' we mean that the couplings are restricted
to a finite range). In general, $\lambda$ may have a Dirac structure,
but in the following we assume its form to be scalar (in Dirac space).
Thus the GWR (\ref{GW2}) with $\alpha =const.$ simplifies to
\begin{equation} \label{GW3}
\lambda (r) = \alpha \sum_{y \in \Z^{d}} [\lambda (y) \lambda (r-y)
- \rho_{\mu}(y) \rho_{\mu}(r-y)] \ .
\end{equation}
For an ultralocal action, the right-hand side extends at least to
twice the range of the left-hand side (it extends even further if
$\alpha$ is momentum dependent) and it is very unlikely that
the latter can achieve cancelations everywhere beyond the range
of $\lambda$, match with the left-hand side at each point inside
that range, and obey the correct normalization.
\footnote{Summation over $r$ leads to the requirement
$\sum_{y} \rho_{\mu}(y)=0$. This is guaranteed for the sensible
assumption that $\rho_{\mu}$ is odd in $y_{\mu}$, which holds
for the FPA in the block average scheme
with an even Dirac scalar $\alpha$.
There, $\rho_{\mu}$ is in addition even in all 
$y_{\nu},\ \nu \neq \mu$, and $\lambda$ is even in all directions.}
All this tends to strongly over-determine the degrees of freedom
in an ultralocal action.
This suggests that in the most local exact 2d and 4d solutions, 
the couplings in $G^{-1}(r)$ decay exponentially in $\vert r\vert$.
\footnote{The simplification \`{a} la eq. (\ref{GW3})
for $\alpha = \pm 1/2$ is solved in $d=1$ by the Wilson action with 
Wilson parameter $r_{W}=\pm 1$.}

Considering eq. (\ref{GW1}), one is tempted invent
new solutions by hand, such as
\begin{equation}
G(p) = \frac{1}{i \bar p_{\mu} \gamma_{\mu} } + \alpha \ , \
G^{-1}(p) = \frac{i \bar p_{\mu} \gamma_{\mu} + \alpha \bar p^{2}}
{1+ \alpha^{2} \bar p ^{2}}
\end{equation}
where $\bar p_{\mu}$ is some lattice version of $p_{\mu}$,
and $\alpha$ is a scalar.
But for instance the obvious example $\bar p_{\mu} = \sin p_{\mu}$, 
which corresponds
in fact to a local action for any real $\alpha$,
suffers from fermion doubling.
Now one has to remove the doublers, maintaining the locality of
$\{ G,\gamma_{5} \}$ (and preferably also of $G^{-1}$),
and it is not trivial to find sensible
alternatives to the fixed point propagator \cite{non-loc}.

However, if we can not achieve ultralocality, in practice a
truncation of the couplings to a short range is needed, which
causes a violation of the GWR. Therefore, what one should aim at
is a truncation that keeps this violation small, and here
the optimization of locality helps. 
The truncation can be carried out in an elegant way
by means of periodic boundary conditions, which keep the normalization
exact. This truncation to couplings in a unit hypercube
-- for the block average FPA with $\alpha =1/2$ --
has been presented in Ref. \cite{Lat96} (``hypercube fermion''), 
and its spectral and thermodynamic properties are drastically improved
compared to the Wilson fermion \cite{Lat96,chem}.
Table \ref{tabtrunc} shows that also the GWR is 
approximated very well by this truncated perfect fermion.

{\footnotesize
\begin{table}
\begin{center}
\begin{tabular}{|c|c|c|}
\hline
$r$ & $\gamma_{5} \{ G^{-1}(r),\gamma_{5} \}$ 
& $\gamma_{5} \sum_{x} G^{-1}(x)\gamma_{5} G^{-1}(r-x)$ \\
\hline
\hline
$d=4$  & & \\
\hline
(0000) & 3.70544 & 3.70544 \\
\hline
(0001) & -0.12152 & -0.11735 \\
\hline
(0011) & -0.06007 & -0.06014 \\
\hline
(0111) & -0.03194 & -0.03356 \\
\hline
(1111) & -0.01685 & -0.01805 \\
\hline
(0002) & 0 & -0.00417 \\
\hline
(0012) & 0 & 0.00020 \\
\hline
(0112) & 0 & 0.00069 \\
\hline
(1112) & 0 & 0.00047 \\
\hline
(0022) & 0 & -0.00034 \\
\hline
(0122) & 0 & -0.00013 \\
\hline
(1122) & 0 & -0.00008 \\
\hline
(0222) & 0 & -0.00006 \\
\hline
(1222) & 0 & -0.00005 \\
\hline
(2222) & 0 & -0.00002 \\
\hline
\hline
$d=2$ & & \\
\hline
(00) &  2.97909 &  2.97909 \\
\hline
(01) & -0.48955 & -0.48631 \\
\hline
(11) & -0.25523 & -0.26035 \\
\hline
(02) &  0  &  -0.00324 \\
\hline
(12) &  0  &  -0.00350 \\
\hline
(22) &  0  &  -0.00188 \\
\hline
\end{tabular}
\end{center}
\caption{The two sides of the Ginsparg-Wilson relation (\ref{GW2})
with $\alpha =1/2$ for the truncated block average FPA 
(``hypercube fermion'') in $d=4$ and in $d=2$. 
For all vectors $r$, which do not occur in the table, both quantities 
vanish. The exact agreement at zero distance is a consequence
of the truncation by periodic boundary conditions.}
\label{tabtrunc}
\end{table}
}
This provides hope that the desired properties related to the remnant
chiral symmetry are realized to a good approximation for the ``hypercube
fermion'' (HF), in particular the absence of additive mass renormalization.
That property is exact for the perfect and for the classically perfect
action \cite{HasRen}. While the index theorem can be considered as
classical, it is especially remarkable that also a quantum
property like the stability of the chiral limit under renormalization
is fulfilled by the classically perfect action.
Intuitively, it appears that in the chiral limit the classically
perfect actions also displays properties of quantum perfection,
and it is plausible that the continuous remnant
chiral symmetry (\ref{trafo})
protects the zero bare mass from renormalization,
similar to the remnant chiral symmetry $U(1) \otimes U(1)$
of the staggered fermions. Here we have the 
additional virtue that the number of flavors is arbitrary (since
doubling is avoided). 

As a further virtue of the classically perfect action
in the chiral limit, there are no exceptional configurations
as shown in Ref. \cite{HLN} (in the sense they are understood there).
However, this paper refers to a fixed point action, which is
defined implicitly by minimization. For practical applications,
we need a very {\em simple} gauging prescription
(involving only a few short lattice paths), and the requirement to
preserve the GWR to a good approximation could guide the
construction of such a good but simple gauging. We comment on
this concept in the Appendix.\\


{\bf II. A new class of solutions of the GWR}\\

A different solution of the GWR arose from
the so-called overlap formalism \cite{NeuGW}. To turn it the
other way round, for any solution of the GWR,
the fermion determinant factorizes in a way compatible with
a vacuum overlap \cite{Narr}.
Also this consideration focused on $\alpha =1/2$, 
and it is interesting that
an entirely different approach singles out the same
$\alpha$ as the optimally local block average fixed point fermion.

We re-derive the solution of Ref. \cite{NeuGW} and reveal
its minimal assumption, which allows for a broad generalization.
First we define
\begin{equation}
V = 1 - G^{-1} \ ,
\end{equation}
and eq. (\ref{GW2}) with $\alpha = 1/2$ is equivalent to
\begin{equation} \label{VV}
\gamma_{5} V \gamma_{5} = V^{-1} \ .
\end{equation}
For the case of one flavor, the solution of Ref. \cite{NeuGW}
reads
\begin{equation} \label{alt}
V = X (X^{\dagger}X)^{-1/2}\ , \ X = 1 - D_{W},
\end{equation}
where $D_{W}$ is the standard Wilson-Dirac operator.

Depending on the gauge background,
there is a danger of a zero eigenvalue of $X$,
so that the expression for $V$ is undefined, but following
Ref. \cite{NeuGW} we assume that this can be ignored
statistically. Note that whenever $V$ is well-defined, then it is
unitary.
Hence all we need for eq. (\ref{VV}) to hold, is the property
\begin{equation}
\gf D_{W} \gf = D_{W}^{\dagger} \ ,
\end{equation}
since this behavior is inherited by $X$ and $V$ (and by $G^{-1}$).
This leaves a lot of freedom for generalizations.
The minimal condition is $\gf X \gf = X^{\dagger}$, but
we stay with the form
\begin{equation}
X = 1 - D
\end{equation}
and note that one may insert many lattice Dirac operators 
for $D$. If we make also here the ansatz
$D(r) = \rho_{\mu}(r)\gamma_{\mu} + \lambda (r)$ -- with the 
normalization mentioned before, and with a scalar form of $\lambda$
-- then this property always holds.

Now the question arises, what generalization is useful.
Locality is a criterion, but even
if $D$ is ultralocal -- as in the case $D=D_{W}$ -- then the decay
of the couplings in $G^{-1}(r)$ is still exponential (in $d>1$).
For $D_{W}$ this can be seen from
\begin{eqnarray}
G^{-1}(p) &=& 1 - \Big[ 1 - i \sin p_{\mu} \gamma_{\mu}
- \frac{r_{W}}{2} \hat p^{2} \Big] \times \\
&& \Big[ 1 + (1-r_{W}) \hat p^{2} + 
\frac{r_{W}^{2}-1}{4} \sum_{\mu} \hat p_{\mu}^{4} + 
\frac{r_{W}^{2}}{2} \sum_{\mu > \nu}
\hat p_{\mu}^{2} \hat p_{\nu}^{2} \Big] ^{-1/2} \ , \nonumber
\end{eqnarray}
which simplifies for the standard Wilson parameter $r_{W}=1$ 
and for $r_{W}=-1$ (the value that the domain wall literature
deals with).
Locality is realized for any choice of $r_{W}$.
\footnote{For $r_{W}=1/d$ the form of $G^{-1}(r)$ 
simplifies to $\delta_{r,0} +$ even-odd-couplings, but the latter
decay rather slowly. I thank M. Peardon for this remark.}

Returning to our general ansatz for $D$, we obtain
\begin{equation} \label{expa}
G^{-1}(p) = 1 - [ 1 - \rho_{\mu}(p) \gamma_{\mu}
- \lambda (p) ] [1 -2 \lambda (p) - \rho^{2}(p) + 
\lambda^{2}(p) ]^{-1/2},
\end{equation}
and together with eq. (\ref{GW3}) this leads to an amusing 
observation:
if we insert an operator $D$, which obeys the GWR itself,
then $X$ becomes unitary, and we end up with
\begin{equation}
G^{-1}=D \ . 
\end{equation}
For example, the inverse fixed point propagator
reproduces itself identically, if we insert it for $D$.
If $D$ violates the GWR, then it gets {\em ``GWR corrected''}.
This correction cannot cure all diseases, however:
if $D$ is non-local or plagued by doubling, then the same
is true for $G^{-1}$. On the other hand, also massive fermions
can be GWR corrected.

As a further criterion, one should aim at small lattice artifacts.
This suggests to insert a choice for $D$, which has good properties
in view of approximate rotation invariance etc., because such
properties are essentially inherited by $G^{-1}$, as the expansion
(\ref{expa}) shows.
Hence, as an experiment we insert the HF
for $D$. Its small GWR violations (see Table \ref{tabtrunc})
are corrected in $G^{-1}$, but an exponential tail of couplings
is added, so that we obtain something similar to the original
fixed point propagator. Of course, this $G^{-1}$ is not fully
perfect, but let's assume that for the aspects of interest only the
GWR matters. Then the action constructed in this way looks
useful, because it is more local than the original FPA
(and much more local than the actions obtained from $D=D_{W}$),
see Table \ref{tabloc}.
Requiring only the GWR means to relax the condition of perfection
a little, and we make use of that by further improving locality.
{\footnotesize
\begin{table}
\begin{center}
\begin{tabular}{|c|c|c|c|c|c|c|}
\hline
& FPA & HF & $r_{W}=0.85$ & $r_{W}=1$ & $r_{W}=1.15$ & $r_{W}=-1$ \\
&     & $\to$ GWR & $\to$ GWR& $\to$ GWR& $\to$ GWR& $\to$ GWR \\
\hline
\hline
$d=4$ &&&&&& \\
\hline
$r_{\rho}$ & 1.635 & 1.519 & 2.688 & 2.530 & 2.537 & 2.424 \\
\hline
$r_{\lambda}$ & 1.187 & 1.109 & 1.819 & 1.708 & 1.810 & 2.784 \\
\hline
\hline
$d=2$ &&&&&& \\
\hline
$r_{\rho}$    & 1.268 & 1.198 & 1.981 & 1.816 & 1.844 & 1.946 \\
\hline
$r_{\lambda}$ & 0.871 & 0.844 & 1.286 & 1.177 & 1.206 & 2.330 \\
\hline
\hline
\end{tabular}
\end{center}
\caption{The locality of the vector and the scalar part
of the FPA, of the GWR corrected version of the ``hypercube
fermion'', and of the GWR corrected Wilson fermion with 
various Wilson parameters $r_{W}$.
(The GWR corrected actions are not normalized, 
because otherwise they violate the GWR again.) 
As a characteristic quantity, we measure
$r_{\rho}^{2} = (\sum_{x} \vert \rho_{\mu}(x) \vert x^{2} ) /
 (\sum_{x} \vert \rho_{\mu}(x) \vert )$, and analogously 
$r_{\lambda}$.}
\label{tabloc}
\end{table}
}
For comparison, the leading
couplings for some of the fermion actions discussed here
are listed in Table \ref{tabcop}.
{\footnotesize
\begin{table}
\begin{center}
\begin{tabular}{|c|c|c|c|c|c|c|}
\hline
& FPA & HF & HF & $r_{W}=1$ & $r_{W}=-1$ & HF $\to$ GWR \\
&     &    & $\to$ GWR & $\to$ GWR & $\to$ GWR & trunc. again\\
\hline
\hline
$d=4$ &&&&& \\
\hline
$\rho_{1}(1000)$ & 0.1400 & {\bf 0.13685} & 0.1363 & 0.1668 & 0.1023 
& {\bf 0.13633} \\
\hline
$\rho_{1}(1100)$ & 0.0314 & {\bf 0.03208} & 0.0318 & 0.0239 & 0.0105 
& {\bf 0.03185} \\
\hline
$\rho_{1}(1110)$ & 0.0096 & {\bf 0.01106} & 0.0110 & 0.0050 & 0.0021 
& {\bf 0.01108} \\
\hline
$\rho_{1}(1111)$ & 0.0033 & {\bf 0.00475} & 0.0048 & 0.0007 & 0.0006 
& {\bf 0.00488} \\
\hline
$\rho_{1}(2000)$ & 0.0040 &      0 & 0.0002 & 0.0238 & 0.0107
& 0 \\
\hline
$\rho_{1}(1200)$ & 0.0015 &      0 &-0.0001 & 0.0053 & 0.0017
& 0 \\
\hline
\hline
$\lambda (0000)$ &  1.8530 &  {\bf 1.85272} &  1.8534 &  1.8334 
& 0.0471 & {\bf 1.85313} \\
\hline
$\lambda (1000)$ & -0.0616 & {\bf -0.06076} & -0.0626 & -0.0560 
& 0.0093 & {\bf -0.06289} \\
\hline
$\lambda (1100)$ & -0.0290 & {\bf -0.03004} & -0.0300 & -0.0241 
& 0.0027 & {\bf -0.03024} \\
\hline
$\lambda (1110)$ & -0.0146 & {\bf -0.01597} & -0.0152 & -0.0115 
& 0.0010 & {\bf -0.01545} \\
\hline
$\lambda (1111)$ & -0.0077 & {\bf -0.00843} & -0.0079 & -0.0057 
& 0.0004 & {\bf -0.00813} \\
\hline
$\lambda (2000)$ &  0.0010 &       0 &  0.0019 & -0.0002 & -0.0038
& 0 \\
\hline
$\lambda (1200)$ & -0.0007 &       0 & -0.0002 & -0.0029 & -0.0005
& 0 \\
\hline
\hline
$d=2$ &&&&& \\
\hline
$\rho_{1}(10)$ &  0.3154 &  {\bf 0.30939} &  0.3089 &  0.3330 
&  0.1680 & {\bf 0.30907} \\
\hline
$\rho_{1}(11)$ &  0.0870 &  {\bf 0.09531} &  0.0953 &  0.0580 
&  0.0273 & {\bf 0.09540} \\
\hline
$\rho_{1}(20)$ &  0.0076 &       0 &  0.0000 &  0.0459 &  0.0288
& 0  \\
\hline
$\rho_{1}(12)$ &  0.0048 &       0 &  0.0002 &  0.0166 &  0.00068
& 0  \\
\hline
$\lambda (00)$ &  1.4903 &  {\bf 1.48954} &  1.4899 &  1.4912 &  
0.0651 & {\bf 1.48955} \\
\hline
$\lambda (10)$ & -0.2441 & {\bf -0.24477} & -0.2464 & -0.2335 &  
0.0173 & {\bf -0.24683} \\
\hline
$\lambda (11)$ & -0.1246 & {\bf -0.12761} & -0.1252 & -0.1167 &  
0.0061 & {\bf -0.12556} \\
\hline
$\lambda (20)$ & -0.0009 &       0 &  0.0015 & -0.0128 & -0.0103
& 0 \\
\hline
$\lambda (12)$ & -0.0032 &       0 & -0.0017 & -0.0108 & -0.0028
& 0 \\
\hline
\hline
\end{tabular}
\end{center}
\caption{The leading couplings of some actions discussed
in the text: the FPA, the ``hypercube fermion'',
the ``GWR corrected hypercube fermion'' and the
``GWR corrected Wilson fermion'' with $r_{W}=1$ (standard)
and $r_{W}=-1$ (standard for domain wall fermions).
All these actions have the symmetries described in footnote 5.
The similarity in the first three cases is
not surprising, but it is amazing that also the fourth example
looks quite similar. For the Wilson fermion we could also
vary $\alpha$ according to eq. (\ref{genalf}).
If we truncate the ``GWR corrected hypercube fermion'' again
periodically, then we reproduce the ``hypercube fermion''
identically, 
hence we truncate in coordinate
space this time and then correct the normalization by hand;
this causes less alteration and yields the last column.
(An iteration of GWR correction and truncation does hardly modify
the couplings any further.)}
\label{tabcop}
\end{table}
}

\begin{figure}[hbt]
\begin{center}
\def\fpsangle{270}
\epsfxsize=75mm
\fpsbox{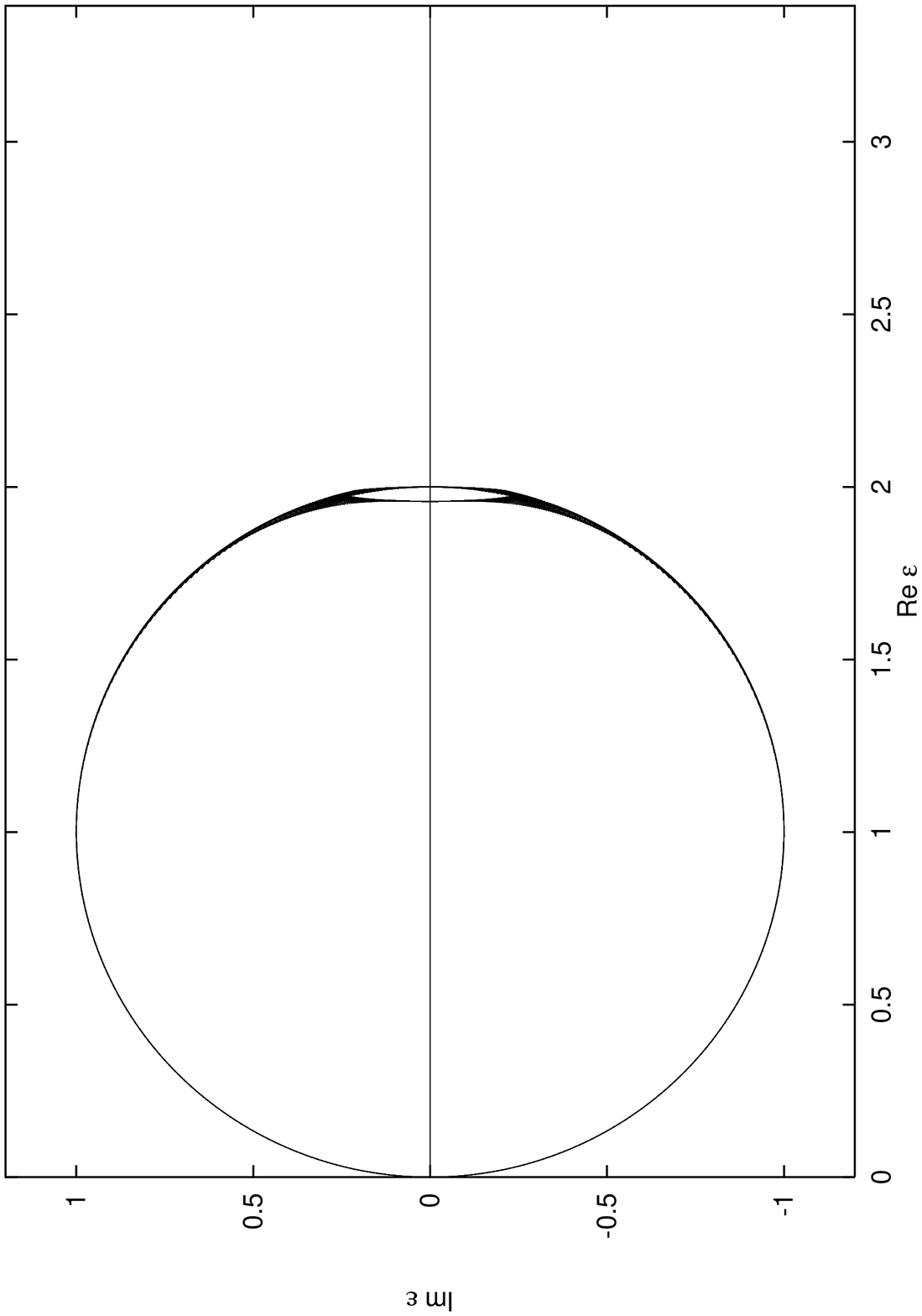}
\end{center}
\vspace{-2mm}
\begin{center}
\vspace{-1cm}
\def\fpsangle{270}
\epsfxsize=75mm
\fpsbox{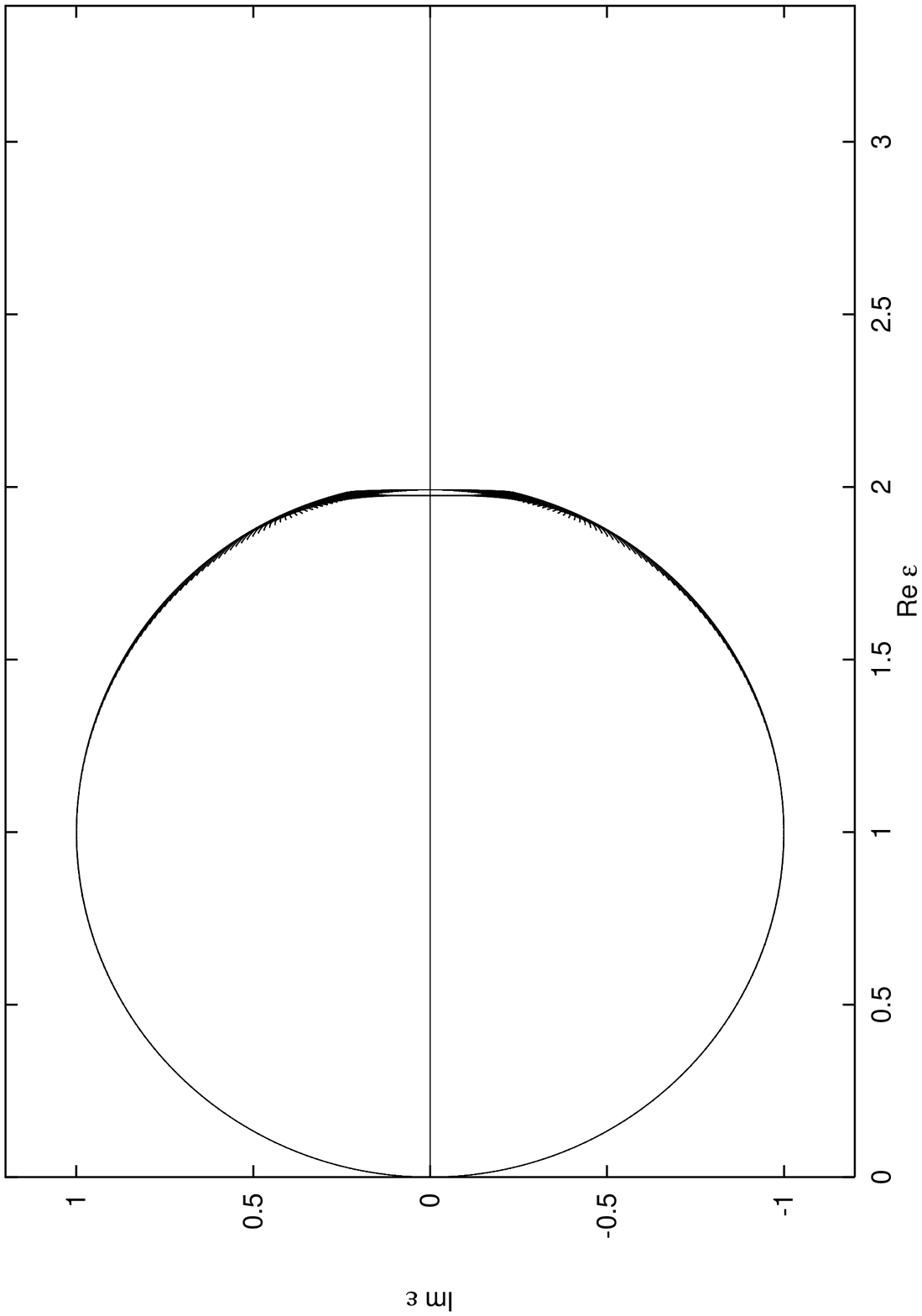}
\end{center}
\vspace{-1cm}
\caption{Spectrum of the ``hypercube fermion'' (on top) and its 
GWR corrected and truncated modification (below) in $d=2$. 
In both cases, the spectra
keep close to the unit circle, hence the artifacts due to truncation
are small. In the most contaminated region around $\vare \simeq 2$
we observe some progress thanks to GWR correction.}
\vspace{-2mm}
\label{spec2d}
\end{figure}

\begin{figure}[hbt]
\def\fpsangle{270}
\epsfxsize=90mm
\fpsbox{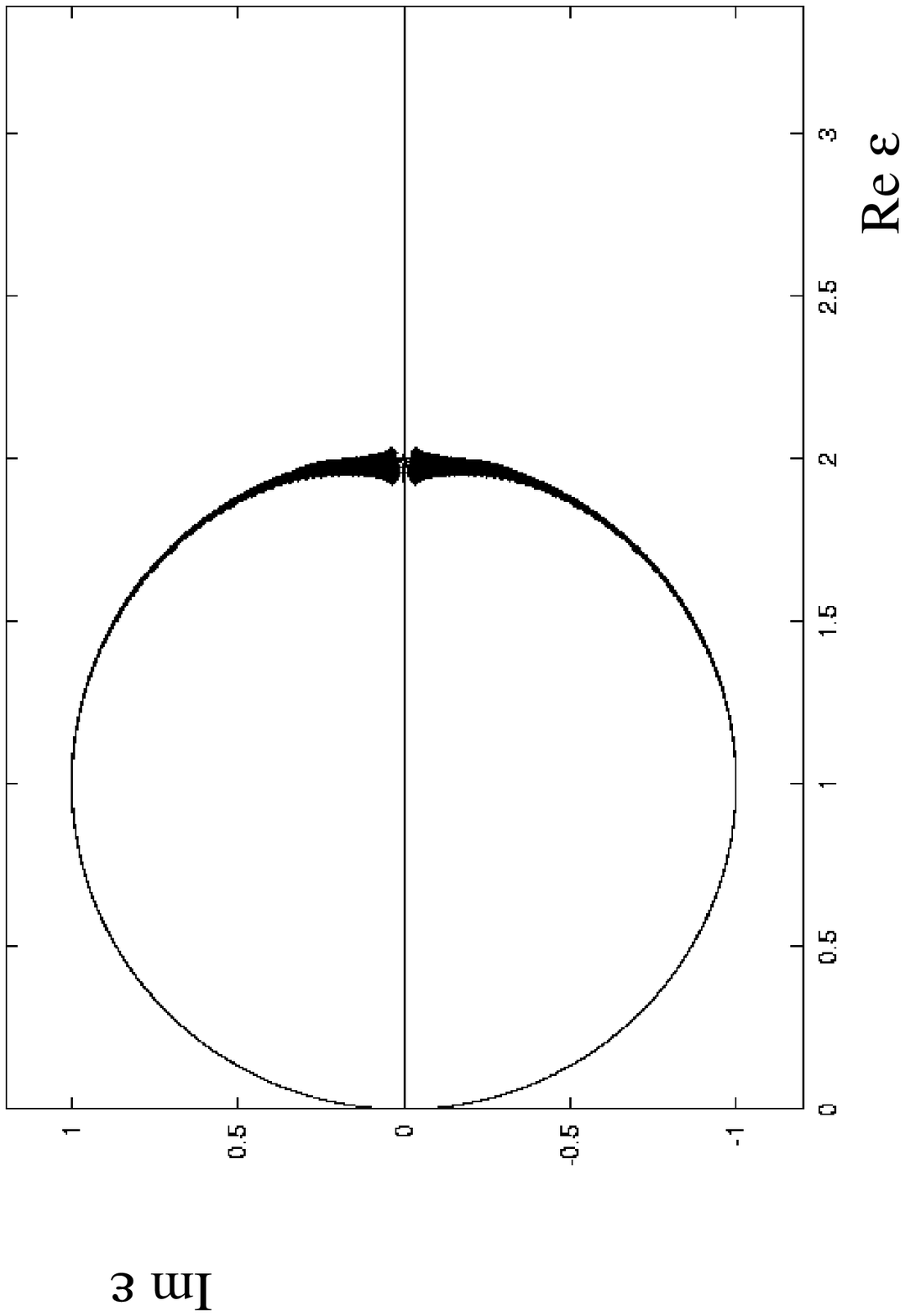}
\vspace{-2mm}
\def\fpsangle{270}
\epsfxsize=90mm
\fpsbox{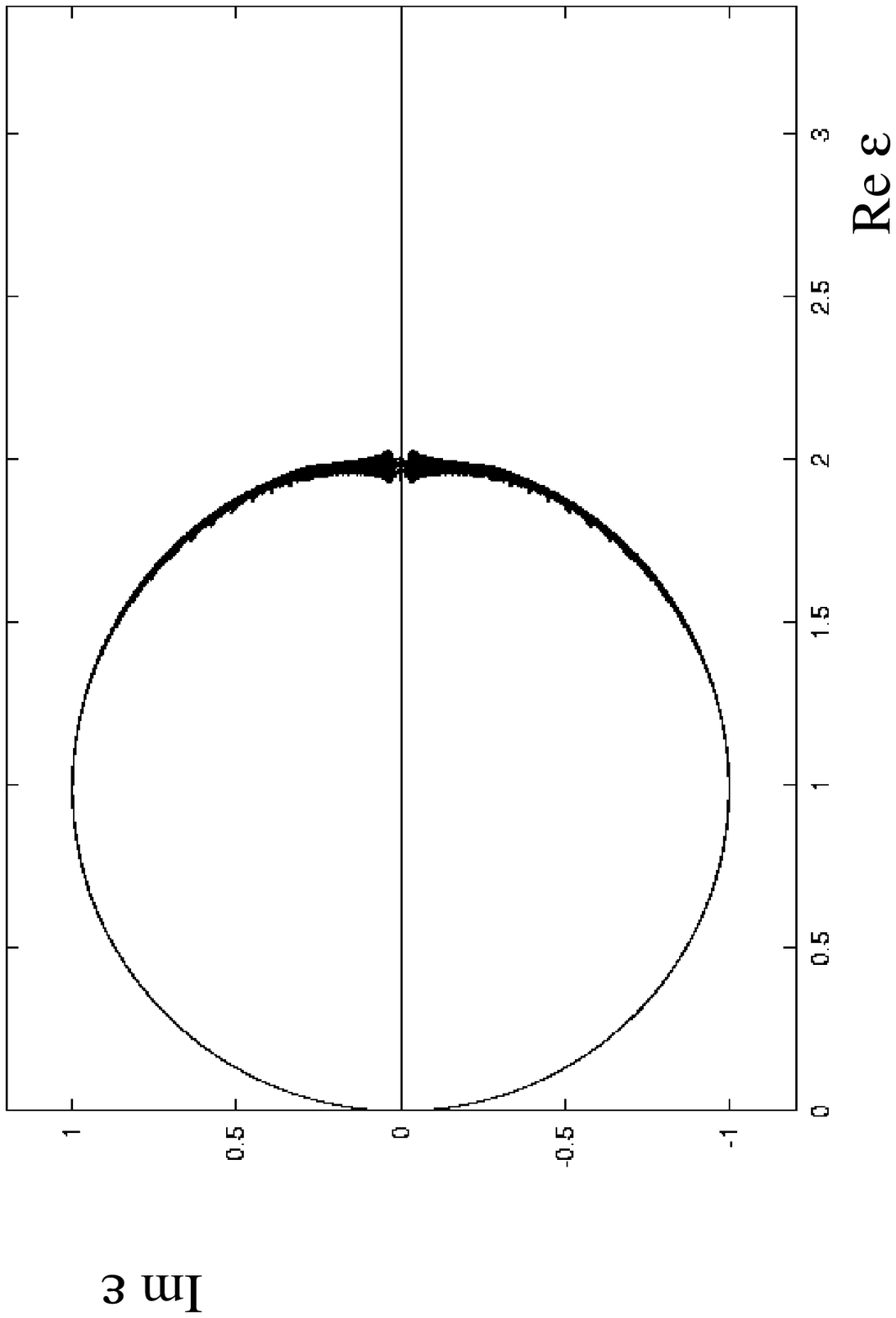}
\vspace{-5mm}
\caption{Spectrum of the ``hypercube fermion'' (on top) and its 
GWR corrected and truncated modification (below) in $d=4$. 
Again, the spectra keep close to the unit circle.
From the shape the progress due to GWR correction is not
easily visible here, but if we measure the mean deviation from the
unit circle, which takes into account the eigenvalue {\em density}, 
then we observe an improvement by 31 percent.}
\vspace{-7mm}
\label{spec4d}
\end{figure}

In order to generalize this class of GWR solutions to any
(non-trivial) local
Dirac scalar $\alpha$, still assuming $\gamma_{5}D \gamma_{5}
= D^{\dagger}$, we write
\begin{equation} \label{genalf}
X = 1 -2\sqrt{\alpha} \ D \ \sqrt{\alpha}\ , \ 
G^{-1} = \frac{1}{2\sqrt{\alpha} } \
\Big[ 1 - X (X^{\dagger}X)^{-1/2} \Big] \ \frac{1}{\sqrt{\alpha}} \ .
\end{equation}
Again, $G^{-1}$ fulfills the GWR, and if also $D$ does so, then
we end up with $G^{-1}=D$, since $X$ is unitary.
The latter further implies that the spectrum of any solution 
$G^{-1}$, at a
specific momentum $p \in B$, is situated on a circle in the
complex plane,
\begin{equation}
\sigma (G^{-1}(p)) \subset \frac{1}{2\alpha (p)}
(1 + e^{i\varphi }) \ , \ \varphi \in [0,2\pi [ .
\end{equation}
The entire spectrum of $G^{-1}$ is located between the
circles given by the minimum and the maximum of $\alpha (p)$.
\footnote{The same geometric structure was obtained in Ref. \cite{HLN}
for the FPA with constant $\alpha$, but iteration of overlapping blocking
functions $\Pi$.}
For our preferred choice $\alpha =1/2$, the entire spectrum
is located on one unit circle. In Figs. \ref{spec2d} and
\ref{spec4d} we illustrate that this property is still
approximated very well for the FPA after truncation. We compare the
HF and its GWR corrected version, which is
truncated again in coordinate space, given in the second and in
the last column of Table \ref{tabcop}. Generally,
the eigenvalues are given in $d=2$ by $\vare_{1,2}(p) =
\lambda (p) \pm i \sqrt{\rho^{2}(p)}$, and in $d=4$ by the roots of
\begin{eqnarray*}
&& \vare^{4}(p)-4\lambda (p) \vare^{3}(p) + [6 \lambda^{2}(p)
+ 2 \rho^{2}(p)] \vare^{2}(p) \\
&& -4\lambda (p) [\lambda^{2}(p)
+ \rho^{2}(p)] \vare (p) + [\lambda^{2}(p)+\rho^{2}(p)]^{2} \ .
\end{eqnarray*}
While the progress due to GWR correction is evident in $d=2$
(Fig. \ref{spec2d}), 
the difference is not easily visible in $d=4$ (Fig. \ref{spec4d}). 
However, if we measure the ``mean
deviation'' from the unit circle by
\begin{equation}
\delta^{2} = \frac{1}{(2\pi )^{d}} \int_{B} dp \ \Big[
\frac{1}{2^{d/2}} \sum_{i=1}^{2^{d/2}} \vert \vare_{i}(p)-1\vert^{2}
\Big] \ ,
\end{equation}
then we observe that $\delta$ decreases from $1.09 \cdot 10^{-2}$
for the HF to $0.75 \cdot 10^{-2}$ for its
GWR corrected and truncated modification.
This shows that the latter solves the GWR to an even better
approximation.

For the free fermion, $p=0$ always yields $\vare_{i}=0$, $i=1\dots
2^{d/2}$, and the whole sector $\vare \approx 0$ is hardly affected
by the truncation (see Figs. 1 and 2).
If we add a gauge interaction, then this sector 
is crucial for the index theorem and for the absence of ``exceptional 
configurations'', hence its good quality is very important.
Truncation effects are rather visible at the opposite sector
$\vare \approx 2$, which corresponds to large momenta.

In a direct application of HFs, it turned out 
that even using a strongly simplified gauging (instead of the 
consistently (classically) perfect incorporation of the gauge 
field), for instance the meson dispersion is drastically
improved \cite{Lat96,Lat97}.
However, this application suffers from a strong mass 
renormalization, which is a problem with respect to practical
issues. In particular, for the minimal gauging by hand along shortest
lattice paths, the ``pion'' mass is renormalized from 0.0 to 3.0
at $\beta = 5$ \cite{Lat96}.
Thus one can hardly trust the couplings in general, and in addition
tuning to the critical bare mass is required.
Its sign is opposite to the sign of $\alpha$, which means that
we are led to a region, which is unfavorable for locality.
Since the truncation effects in the free fermionic FPA are small,
we expect that this renormalization could be almost avoided 
by using a quasi
(classically) perfect gauge interaction, which is, however,
very difficult to realize in $d=4$. 
If we gauge by hand, trying to suppress the additive mass
renormalization (as in Ref. \cite{TdG}), then one should use the
most local free fermion couplings, and here the ``GWR corrected 
hypercube fermion'' is in business. 
For the idea to use the GWR as a guide-line
to construct a better gauging by hand, we refer again to the appendix.\\

{\bf III. Improved domain wall fermions} \\

So far we have considered the construction of a potentially
interesting GWR corrected propagator for the direct use.
However, the way people think about simulating overlap
fermions uses directly $D$ in the above language. 
In that context, the quality of the locality of $G^{-1}$ is not 
directly relevant, hence  one could choose $D=D_{W}$
and then gauging is not problematic either.
On the other hand, one runs into some trouble with `exceptional 
configurations' (now in the sense that $X$ has a zero eigenvalue).
\footnote{The extent of this danger depends on $r_{W}$: for 
$r_{W} \leq 1$ the smallest eigenvalue 
of $X^{\dagger}X$ in the free case is 1, 
but for larger $r_{W}$ it decreases down to a minimum
of 0.288 at $r_{W}=6.4$, before it rises again to 1.}
More importantly, an extra dimension has to be introduced, which
can be interpreted as a fifth direction with $L_{s}$ sites, or simply
as $L_{s}$ flavors. The meaning of the square root, that the fermion 
determinant is divided by, is to subtract the contributions of
heavy modes, generated by the extra dimension. In practice one
has to work with auxiliary boson fields \cite{Vra}.

For practical purposes, the domain wall fermion formulation
by Y. Shamir seems most suitable \cite{Shamir}, hence we want to 
discuss possible improvements in that framework.
In terms of vector-like
gauge theories, first simulations have been performed
for the Schwinger model \cite{Jaster,Vranas}
and for QCD \cite{Blum}. Again, an extra dimension is required,
and the limit $L_{s}\to \infty$ has attractive features:
absence of additive mass renormalization (supported by a one loop
calculation \cite{Aoki}), for the measurement
of matrix elements no fine tuning is needed either
\cite{Blum}, and there are only $O(a^{2})$ artifacts
\footnote{This is related to the chiral symmetry at $L_{s}\to \infty$,
which rules out all $O(a)$ operators. Again, there is a certain 
similarity with staggered fermions, but here we keep the whole
$SU(N)\otimes SU(N)$ symmetry, hence the correct number of
Goldstone bosons is involved.}.
In practice $L_{s}$ must be finite, so that
these properties are not exact anymore, but one expects
for instance the $O(a)$ artifacts to be suppressed exponentially
with $L_{s}$. Also the chiral limit appears to be stable to a
decent approximation down to $L_{s} \simeq 10$ \cite{Blum}.
Questions of the $L_{s}$ truncation have been discussed
on the theoretical level in Refs. \cite{Shamir,Trunc}.
We note that Shamir's formulation allows to work with $L_{s}/2$
only, in contrast to Kaplan's original proposal \cite{Kaplan}.

The action, which has been used so far, has the following
form \cite{FuSha}: 
in Euclidean space -- at a fixed extra coordinate $s$ --
it is the usual Wilson action
with gauge fields on the links, with $r_{W}=-1$ and a mass
$M$. In the extra dimension, it is the free massless Wilson 
action with $r_{W,s}=-1$ (the kinetic term with $\gamma_{5}$), 
and with one additional term
\begin{equation}
D^{add}_{s,s'} = \frac{\sigma}{2} \ [ (1-\gamma_{5}) \delta_{s,0}
\delta_{s',L_{s}-1} +  (1+\gamma_{5}) \delta_{s',0}
\delta_{s,L_{s}-1}]
\end{equation}
(we refer to one flavor).
The bare quark mass is $m_{q}= M(M-2)(1+\sigma )$,
but due to renormalization, there is a tuning problem for $M$.
On the other hand, one expects (and observes reasonably well
\cite{Blum}) $m_{\pi}^{2} \propto (1-\sigma)$.

Now we distinguish three possible strategies for the improvement
of this type of domain wall fermions:

{\bf 1)} The observation that certain artifacts of $D_{W}$,
for instance in the rotational invariance,
essentially persist,
motivates the use of the HF -- or of its GWR corrected and
truncated modification -- in the 4d (or 2d) Euclidean space.
Then for example a simplified gauging
would be less problematic,
because the additive mass renormalization and other diseases
are suppressed by $L_{s}$. 
Still, such a simulation would be tedious,
but so is the use of a properly gauged HF.
If this works, then one cumulates all sort of advantages:
very small lattice artifacts (doubly suppressed from the
HF and from $L_{s}$), a small additive mass 
renormalization, an arbitrary number of flavors and the correct
number of Goldstone bosons.
It remains to be checked if the left- and the right-handed fermions 
can still be separated, see below.

{\bf 2)} Instead one could improve in  the 5-direction.
If we assume $L_{s}$ periodicity, then
its inverse propagator in the above Wilson-like form reads
\begin{equation}
D_{W,5}^{-1}(p_{5}) = i \sin p_{5} \gamma_{5} + \frac{r_{W,s}}{2} 
\hat p_{5}^{2} + \frac{\sigma}{L_{s}} \exp (i L_{s} p_{5} \gamma_{5}) \ .
\end{equation}
If we block in the 5 direction only, the corresponding
``FPA'' is given by
\footnote{``FPA'' needs inverted commas, because we are
at finite $L_{s}$, but let's assume that we start at
a huge size $n^{k}L_{s}$, then perform $k$ RGTs with block factor $n$,
so we end up in $L_{s}$. In the limit $n^{k} \to \infty $ we
obtain the expression (\ref{D5}).}
\begin{equation} \label{D5}
D_{5}(p_{5}) = \sum_{l_{5} \in \Z} 
\frac{\Pi_{5}^{2}(p_{5}+2\pi l_{5})}
{(p_{5}+2\pi l_{5}) (1+\sigma ) \gamma_{5} + \sigma /L_{s} }
+ \alpha_{5} \ ,
\end{equation}
where $\Pi_{5}$ is the blocking prescription in the 5-direction.
If we choose again the block average, and we optimize
$\alpha_{5}$ for locality, then we arrive at
\begin{eqnarray}
D_{5}^{-1}(p_{5}) &=& (1+ \sigma ) \Big( \frac{u}{\hat u} \Big)^{2}
\Big[ i \sin p_{5} \ \gamma_{5} \pm (\frac{1}{2} \hat p_{5}^{2} 
+ \hat u ) \Big] \\
u &=& \frac{\sigma}{(1+\sigma) L_{s}} \ , \ \hat u = e^{u}-1 \ , \
p_{5} = \frac{2\pi}{L_{s}} j , \ j = 0 \dots L_{s}-1. \nonumber
\end{eqnarray}
In a sense, there in not much to improve in one dimension,
since the Wilson action (with $ r_{W} =\pm 1$) is already perfect;
we have just incorporated $\sigma$ and $L_{s}$.
However, the $s$ dependent defect, which is needed to separate
the chiral fermions, has disappeared. A way to avoid this
is the use of {\em fixed} boundary conditions in the $s$ direction,
similar to Ref. \cite{Gato}.
This study is in progress.

{\bf 3)} Finally, one could improve directly the entire 5d action,
allowing for diagonal couplings with respect to all directions.
Since there are no gauge fields on the 5-links, this is perhaps
not as disastrous as it first appears.
With respect to gauging, there is no additional problem
compared to the case 1), and we hope to suppress all discretization
artifacts at once. Here, a conspiracy of $M$, $\sigma$ and $L_{s}$
affects the couplings in all directions.

Also in this case, $L_{s}$ periodic boundary conditions destroy the
defect in the $s$ direction 
\footnote{This happens even if we split $\alpha^{-1}$ into
$[\alpha^{-1}_{L}(1-\gamma_{5}) + \alpha^{-1}_{R}(1+\gamma_{5})]/2$,
possibly combined with chirally specific functions $\Pi_{L}$,
$\Pi_{R}$.}, hence a more sophisticated treatment is needed.\\

The defect in the $s$ direction is needed to keep the
chiral fermions apart, so we focus on variant {\bf 1)}, which
leaves the couplings between the 4d layers untouched.
Since in general a mass parameter $M$ appears in the physical
subspace, we insert a massive HF.
This is constructed from the perfect action at mass $M$,
again by truncation with periodic boundary conditions \cite{Lat96}.
There are two choices for $\alpha$ which restrict the 1d action to 
nearest neighbors, $\alpha = \alpha_{M} = (e^{M}-M-1)/M^{2} >0$,
and $\alpha = - \alpha_{-M}$. In higher dimensions, the first (second)
option optimizes the locality for positive (negative) $M$.
\footnote{Generally, the changes $M \to -M$ and $\alpha \to -\alpha$
imply $\rho_{\mu} \to \rho_{\mu}$, $\lambda \to - \lambda$.}
In the following, we choose $\alpha = \alpha_{M}$.
As a sound basis, the hypercubic couplings for typical masses,
which have been used in simulations, $M=1.7$ in $d=4$ 
(the approximate (quenched) critical value
at $\beta = 6$ \cite{Blum}) and
$M=0.9$ in $d=2$ \cite{Jaster,Vranas}, 
are given in Table \ref{tabhf}. Note, however,
that the renormalization with an improved action could
lead to a different critical value of $M$.
{\footnotesize
\begin{table}
\begin{center}
\begin{tabular}{|c|c|c|}
\hline
$r$ & $\rho_{1}(r)$ & $\lambda (r)$ \\
\hline
\hline
& $d=4$ & $M=1.7$ \\
\hline
(0000) &       0 &  0.95860 \\
\hline
(1000) & 0.02606 & -0.01613 \\
\hline
(1100) & 0.00468 & -0.00476 \\
\hline
(1110) & 0.00123 & -0.00178 \\
\hline
(1111) & 0.00041 & -0.00076 \\
\hline
\hline
 & $d=2$ & $M=0.9$ \\
\hline
(00) &       0 &  1.14008 \\
\hline
(10) & 0.14008 & -0.10247 \\
\hline
(11) & 0.03278 & -0.04382 \\
\hline
\hline
\end{tabular}
\end{center}
\caption{The truncated perfect ``hypercube fermion''
at the masses, which have been used in domain wall fermion
simulations: $M=1.7$ in $d=4$, and $M=0.9$ in $d=2$.}
\label{tabhf}
\end{table}
}

Finally, we want to show that it is indeed possible in this
framework to separate the chiral modes. To be able to do so
analytically, we now let $L_{s}\to \infty$, and we switch to Kaplan's
version, with an ordinary Wilson action in the $s$ direction 
plus a mass-like term
$m(s)= m\cdot {\rm sign} (s)$, ${\rm sign}(s)$ = 1 for $s>0$,
$-1$ for $s<0$, and 0 at $s=0$. Hence the free fermion action reads
\begin{eqnarray}
S[\bar \Psi , \Psi ] & \!\! =&\!\!\!\! \sum_{x,r,s,s'} \bar \Psi_{x,s}
\Big[ [ \rho_{\mu}(r) \gamma_{\mu} + \lambda (r)] \delta_{s,s'} 
+ [ \frac{1}{2} (\delta_{s,s'+1} - \delta_{s+1,s'}) \gamma_{5} 
+ m(s) \nonumber \\
&&
+ \frac{r_{W,s}}{2} (\delta_{s,s'+1}+\delta_{s+1,s'} -2\delta_{s,s'})]
\delta_{r,0} \Big] \Psi _{x+r,s'} \ ,
\end{eqnarray}
where $\rho_{\mu}$,
$\lambda$ refer to the HF at some mass $M$.
We now follow the procedure of Ref.~\cite{JaSchma} to look for
solutions of the type \ $\Psi_{R,L} = e^{ipx} \Phi_{s} 
u_{R,L}$ , where $\gamma_{5}u_{R} = u_{R}$, $\gamma_{5}u_{L} = - u_{L}$.
We require the inverse propagator $G^{-1}(p,s)$
to reduce to a chirally invariant $G^{-1}(p)$, 
\footnote{Here we mean full chiral invariance, not just the GWR.
Hence $G^{-1}(p)$ is not a truncated perfect Dirac operator,
but just its vector part $i\rho_{\mu}(p)\gamma_{\mu}$, which
-- in its isolated form -- would imply fermionic doubling.}
which amounts to the condition
\begin{equation}
\Big[ \frac{1}{2} (\delta_{s,s'+1}-\delta_{s+1,s'})\gamma_{5}
+m(s) - \frac{r_{W,s}}{2} (\delta_{s,s'+1}+\delta_{s+1,s'}-
2\delta_{s,s'}) + \lambda (p) \Big] \Phi_{s} u_{R,L} = 0 .
\end{equation}
Following Kaplan \cite{Kaplan} we make the ansatz $\Phi_{s+1}
= z \Phi_{s}$ for $s,s+1 \neq 0$.
The solutions for $z$ are
\begin{equation}
z_{R,L} = \frac{U \pm \sqrt{U^{2}-r^{2}_{W,s}+1}}{r_{W,s} \mp 1} \ , \
U = m(s) + \lambda (p) +r_{W,s} \ .
\end{equation}
The upper (lower) sign in the denominator refers to $z_{R}$ ($z_{L}$),
and in both cases the two signs in the numerator are possible.
As an example, in the limit $r_{W,s} \to 1$ there is only
one finite solution: $z_{R}=1/U$. Normalizability requires
\begin{eqnarray}
\vert z_{R}(s>0)\vert &=& \vert \frac{1}{1+m+\lambda (p)}\vert < 1
\ \rightarrow \ \lambda (p) > -m \ \ {\rm or} \ \ \lambda (p) < -(2+m)
\nonumber \\
\vert z_{R}(s<0)\vert &=& \vert \frac{1}{1-m+\lambda (p)}\vert > 1
\ \rightarrow \ (m-2)< \lambda (p) < m \ .
\end{eqnarray}
The allowed region for $\lambda (p)$ is shown as shaded areas in
Fig. \ref{figlam} on top. Below
we show the interval of values that $\lambda (p)$ actually takes,
depending on $M$ (for $\alpha = \alpha_{M}$). 
The minimum is at $\lambda (p=0)=
M^{2}/(e^{M}-1)$, and we see that there are many possible combinations
of $m$ and $M$ which yield a single right-handed fermion. In some
cases the solution collapses at some ``critical momenta'', in others
it extends all over the Brillouin zone. It is particularly
favorable to choose a rather large parameter $M$, so that 4d locality 
is excellent and -- as a related property --
$\lambda (p)$ is confined to a narrow interval for all momenta.
Then, for instance $m\approx 1 \dots 2$ guarantees the solution 
to exist at any momentum.

\begin{figure}[hbt]
\begin{center}
\def\fpsangle{270}
\epsfxsize=73mm
\fpsbox{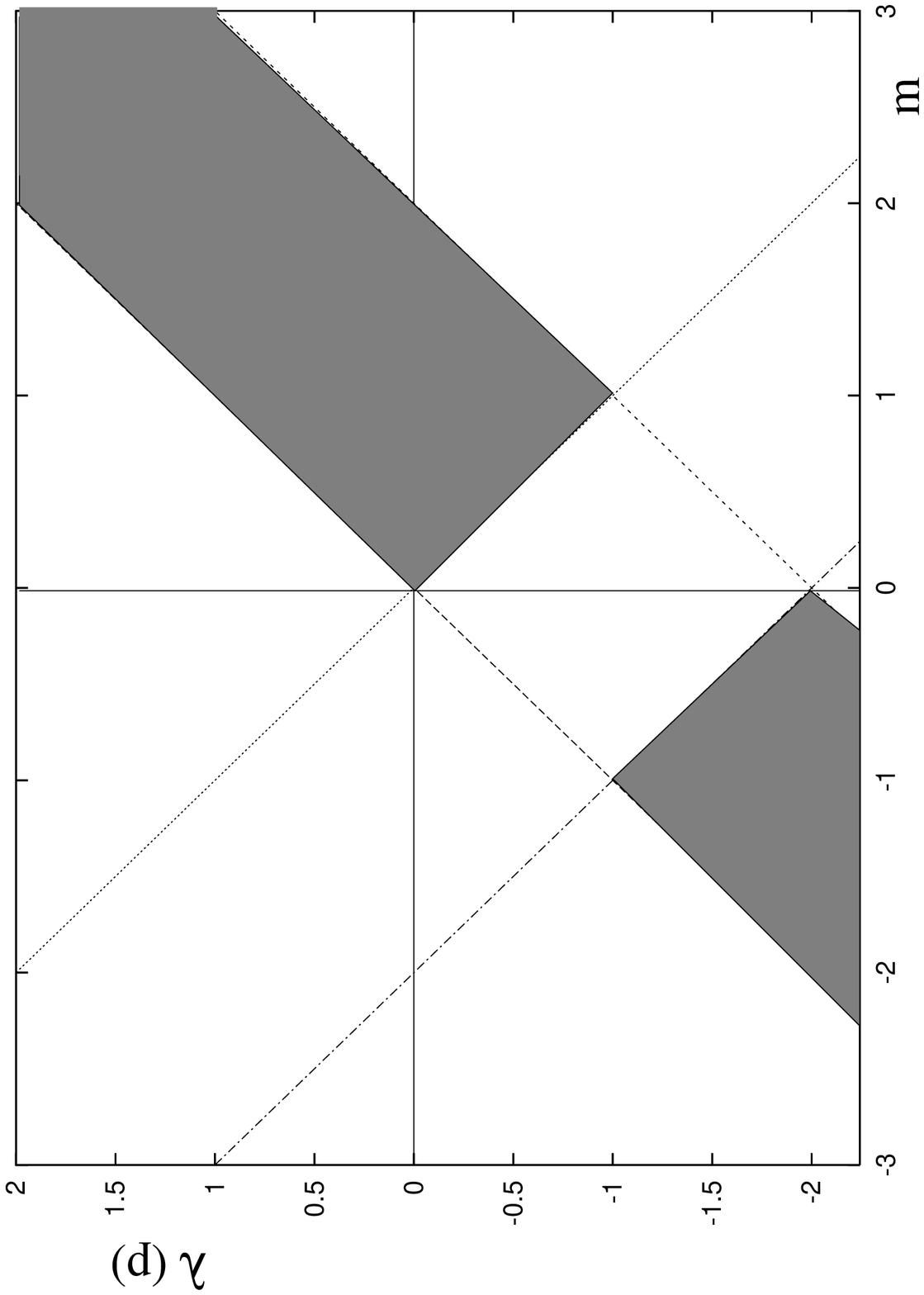}
\end{center}
\vspace{-8mm}
\end{figure}
\begin{figure}[hbt]
\begin{center}
\vspace{-1cm}
\def\fpsangle{0}
\epsfxsize=100mm
\epsfbox{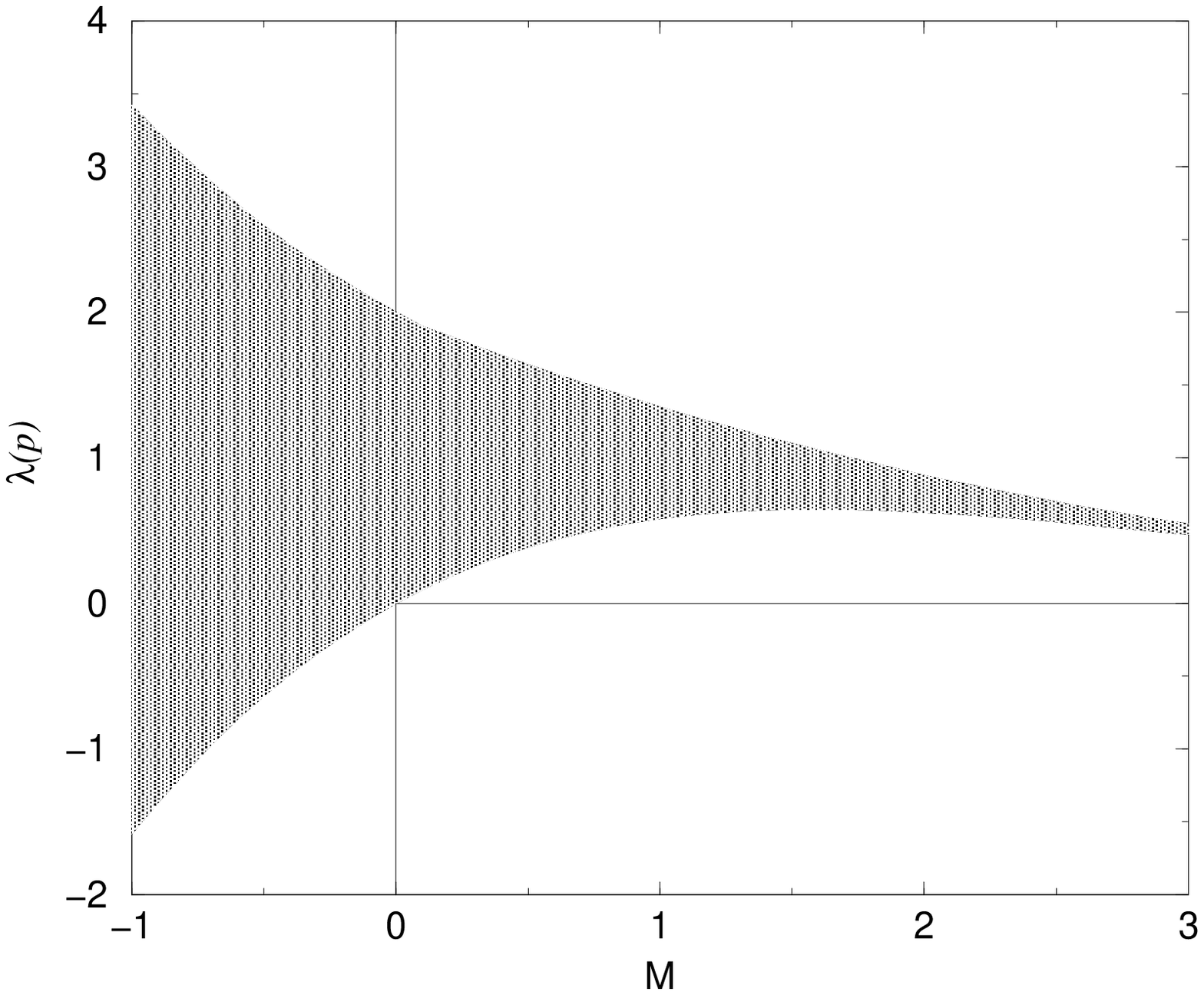}
\end{center}
\vspace{-1cm}
\caption{On top, the shaded regions are the allowed areas for
$\lambda (p)$ providing a right-handed Kaplan-type fermion for
$r_{W,s}=1$ and $m(s) = m \cdot {\rm sign}(s)$.
Below, we show the interval where $\lambda (p)$ takes
its values for a given parameter $M$ and $\alpha = \alpha_{M}$.
(For $\alpha = - \alpha_{-M}$ the shaded area is mirrored at the
origin.)}
\label{figlam}
\end{figure}

In the limit $r_{W,s} \to -1$ the situation is analogous, but there one
deals with a left-handed fermion.\\

{\bf IV. Conclusions} \\

We have revisited the fixed point action approach to the formulation
of lattice chiral fermions, emphasizing in particular the blocking
from the continuum and the r\^{o}le of the Ginsparg-Wilson relation.
We also consider the artifacts in the remnant chiral symmetry 
in short ranged truncated fixed point actions. 
Based on a formula used in the overlap
formalism, we found a method how a lattice Dirac operator
can be ``GWR corrected''. This leads to a large class of new
solutions, and it allowed us to further optimized their locality,
which pays off in reduced spectral artifacts after truncation.
Along these lines, we also arrived at a formulation of
improved domain wall fermions. 
We discussed practical issues in view of their
application in simulations of vector-like theories. 
The improvement could, for instance, reduce the extent $L_{s}$
in the fifth direction, which is needed for a number of favorable
properties. Finally, we showed
that the mechanism to separate the chiral fermions still
works in our improved formulation.\\

{\em Acknowledgment} I am indebted to S. Chandrasekharan,
F. Niedermayer, K. Orginos, M. Peardon and U.-J. Wiese 
for interesting comments.\\


{\bf Appendix} \\

We know that FPAs obey the GWR, which implies plenty of attractive 
properties, but we also know that such actions are extremely hard 
to implement. What we need is a good but very simple gauging of our
HF, and here we want to comment briefly on the possibility to use
the GWR for the construction of such a gauging.

We denote the lattice Dirac operator by $D$ and refer to $\alpha
=1/2$, so that the GWR can be written as
\begin{equation}
D(r) + \bar D(r) = \sum_{x} \bar D (x) D(r-x) \ , \quad
\bar D = \gamma_{5}D\gamma_{5} \ .
\end{equation}
Let $D(r)$ represent the HF in the free case, which obeys the GWR
to a good accuracy (see Table \ref{tabtrunc}).
The task is now to find a gauging which preserves
this property in the interacting case. 
This is rather involved, because now the GWR is supposed to
hold approximately for each lattice path. (The situation is
even worse in cases where $\alpha \neq const.$).

As a simple example, we consider the gauging by the average of the
shortest lattice paths only, and we discuss the case $d=2$.
Regarding condition (\ref{GW3}), the case $r=(00)$ remains exact,
and for $r=(11),\ (21)$ and (22) the combination of couplings,
which nearly matches in the free case, acts on all paths involved,
hence the GWR is still approximated well. For $r=(20)$ that
combination of couplings
splits into coefficients for the paths of length 2 and 4, but
all those coefficients are small, so the situation is not much
worse than in the free case. The main problem is the case
$r=(10)$, which amounts to
\begin{equation}
-0.490 \ [ \ link \ ] = -0.608 \ [ \ link \ ] + 0.061 \ 
[ \ sum~over~staples \ ] \ .
\end{equation}
Here the matching path by path does not work well, and this provides
some insight into the limitation of the minimal gauging by shortest
lattice paths.
One should now correct for that by inserting fat links, a clover
term etc. and tuning its coefficients. This work --
and the extension to $d=4$ -- is in preparation.

At this point, we just add a remark on how to simplify this
task. If we generally
define $X=1-D$, $\bar X = \gamma_{5}X\gamma_{5}$,
then the GWR for $\alpha = 1/2$ reads
\begin{equation}
\bar X X =1 \ .
\end{equation}
If $D$ is linear in the $\gamma_{\mu}$'s,
$D = \rho_{\mu}\gamma_{\mu} + \lambda$,
then the GWR is identical to the requirement that $X$ be unitary.
We insert again the HF for $D$ and introduce a new hypercubic variable
$\tilde \lambda = 1 - \lambda$, so the GWR takes the form
\begin{equation}
\sum_{x}[ -\rho_{\mu}(x) \rho_{\mu}(r-x) + \tilde \lambda (x)
\tilde \lambda (r-x) ] = \delta_{r,0} \ .
\end{equation}
Again this holds exactly at $r=0$ for any reasonable gauging,
but $r\neq 0$ is complicated. Of course, we cannot avoid
the inconvenient quadratic term by these re-definitions,
but at least the rest has been optimally simplified.


\begin{thebibliography}{40}
\bibitem{non-loc} Since the lattice Dirac operator 
$D_{\chi}(p)=i\bar p_{\mu}\gamma_{\mu}$ 
is chirally invariant, it must violate
an assumption of the Nielsen-Ninomiya theorem. Most violations --
such as doubling, non-Hermiticity, non-technicality, non lattice
translation invariance, or non-locality in the sense that
$D_{\chi}(p)$ performs finite gaps -- are still present
in $G^{-1}(p)$ and therefore unfavorable.
However, we can allow for a non-locality in the sense that
$D_{\chi}(p)$ has divergences, and still obtain a local
$G^{-1}(p)$. This is exactly the mechanism to proceed from a
$\delta$ function FPA to a local FPA.
\end{thebibliography}
\end{document}